\documentclass[english,aps,pra,reprint,showpacs]{revtex4-1}
\usepackage[T1]{fontenc}
\usepackage[latin9]{inputenc}
\usepackage{amstext,amsmath}
\usepackage{amssymb}
\usepackage{graphicx}
\usepackage{color}
\usepackage{esint}

\makeatletter
\usepackage{dsfont}
\usepackage{braket}
\usepackage{appendix}

\makeatother

\usepackage{babel}

\newcommand{\matel}[3]{{\left\langle \vphantom{#1 #2 #3} #1 \,\right\vert
\left.
 \hspace{-0.15em} \vphantom{#1 #2 #3} #2 \,\right\vert \left.
 \hspace{-0.15em} \vphantom{#1 #2 #3} #3\right\rangle}}

\begin{document}

\title{Collisional decoherence of polar molecules}

\author{Kai Walter}
\author{Benjamin A. Stickler}
\author{Klaus Hornberger}

\affiliation{Faculty of Physics, University of Duisburg-Essen, Lotharstra\ss e 1, 47048 Duisburg,
Germany}

\pacs{03.65.Yz,34.50.-s,03.75.-b}

\begin{abstract}


The quantum state of motion of a large and rotating polar molecule can lose coherence through the collisions with gas atoms. We show how the associated quantum master equation for the center-of-mass can be expressed in terms of the orientationally averaged differential and total scattering cross sections, for which we provide approximate analytic expressions. The master
equation is then utilized to quantify collisional decoherence in a 
interference experiment with polar molecules.

\end{abstract}

\maketitle

\section{Introduction}

Coherence experiments with molecules and molecular clusters test the quantum
superposition principle and allow measurement of molecular vacuum
properties with unprecedented accuracy
\cite{Schollkopf1994,Arndt1999,brezger2002,Milner2014}. Performing such experiments with organic molecules or biomolecules 
\cite{eibenberger2011,Gerlich2011,Doerre2014,GeyerBio2016}, which are
predominantly polar, raises the question of how their quantum coherence is affected by collisions with a background gas.
It is the aim of the current article to clarify the
influence of molecular dipole moments on collisional decoherence due
to scattering with gas atoms.

Collisional decoherence is ubiquitous, its relevance reaching well
beyond matter-wave experiments
\cite{schlosshauer,joos2003,zurek2003,Trost2009,gonzalo2011stabilization,coles2012consistent}. While
it is still unclear whether the quantum
superposition principle is valid at all scales \cite{adler2004quantum,bassi2013models,Arndt2014Testing}, the effect of
decoherence due to scattering with environmental particles provides an
experimentally tested quantum description of the appearance of
classical dynamics \cite{Hornberger2003,Milner2014,tenney2016collisional}.

In the case of a spherical, polarizable particle, the quantum linear Boltzmann equation gives a
Markovian, nonperturbative microscopic description of the particle's
motion through a thermal environment
\cite{Hornberger2006,Vacchini2009}. If the molecule is much heavier than the gas particle, this master equation can be expressed in terms
of a scattering rate and a decoherence function. These quantities are
determined by the total and the differential scattering cross section
of a single collision, as calculated using standard
scattering theory \cite{taylor}.

In order to formulate the theory of collisional decoherence for
nonspherical particles, one must take the molecule's center of mass
as well as its orientational degrees of freedom into account. In
general, this implies solving the full inelastic scattering problem to
obtain the scattering amplitudes required for the master
equation \cite{PhysRevA.82.042111,hemming2010collisional,Trost2009}. Yet, if the molecule is
sufficiently massive and approximately static during a
single collision, the scattering event is effectively
elastic \cite{Bennewitz1964}. We will see that the decoherence
function and the scattering rate then depend on the orientationally
averaged cross sections. Here, we can draw on the
orientation-dependent cross sections, that were calculated and
measured in various molecular collision experiments
\cite{Bennewitz1964,Bernstein1964,ChildBook,PaulyDiffSig}. We extend
these works and derive the resulting decoherence function and
scattering rate. It is shown that both quantities can be measured by
placing a collision chamber in a matter-wave interferometer. Comparison of the scattering rate of two different polar
molecules provides a nonspectroscopic means to measure their relative electric dipole moment in the
gas phase.

We note that our analytic expressions for the 
orientation-dependent total and differential cross sections may also become relevant for future
experiments where quantum coherence plays no role. This includes buffer gas cooling of large polar
molecules \cite{patterson2010cooling,hutzler2012buffer} and
dedicated collision experiments with molecular beams 
\cite{vogels2014high}.
Moreover, they can be used to assess the influence of background gases in molecular  beam
deflection measurements \cite{miyajima2008stern, 1.3610390,pauly2013atom}.

The paper is structured as follows: In Sec.~\ref{sec:colldec} we
present the master equation for the center-of-mass degrees of freedom
of a slowly rotating molecule in a homogeneous gas environment.
Section~\ref{sec:crosssec} considers the scattering between a
non-spherical particle and a single gas particle in order to obtain
approximate expressions for the total as well as the differential
scattering cross section, and in Sec.~\ref{sec:decscatt} we  derive
the decoherence function and the scattering rate. In
Sec.~\ref{sec:application} collisional decoherence in a typical
far-field matter-wave setup is studied as an application. We conclude
in Sec.~\ref{sec:conc}.

\section{Master Equation for the center-of-mass motion\label{sec:colldec}}

We consider a rigid top molecule of mass $M$ which enters with velocity ${\bf v}_{\rm M}$ a chamber
filled with a mono-atomic gas at temperature $T$. The molecular degrees of freedom (DOFs) are its
center-of-mass (c.m.) position $\mathbf{R}$ and its orientation $\Omega$, specified for instance by
the Euler angles in the $z$-$y'$-$z''$-convention \cite{Sakurai,Brink}. Since the gas is at thermal
equilibrium, the state of the gas $\rho_{\rm g}$ is diagonal in the momentum basis and its diagonal
elements are given by the Boltzmann distribution $\mu( \vert {\bf p} \vert )$.

For isotropic interaction potentials a Markovian master equation can
be derived by means of the so-called monitoring approach
\cite{Hornberger2007,Hornberger2008}. The resulting
equation is then characterized by the rate operator $\mathsf{\Gamma}$
as well as by the scattering operator $\mathsf{S}$, which describes
the modification of the composite molecule-gas state $\rho_{\rm tot}$
by a single scattering event, $\rho_{\mathrm{tot}}\to \rho_{\rm tot}'
= \mathsf{S}\rho_{\rm tot}\mathsf{S}^{\dagger}$.

In the case of non-spherical molecules, the interaction potential is
anisotropic and the orientation state must be taken into account. This
leads in general to a multi-channel scattering problem
\cite{PhysRevA.82.042111} where the Schr{\"o}dinger equation must be
solved for arbitrary initial and final rotation states of the molecule
\cite{Arthurs540}. However, when considering the room-temperature
collision between a multi-atomic molecule and a fast single atom, the
rotational period is typically much smaller than the collision time
and the molecule can be assumed to be rotationally static during the
scattering process (sudden approximation) \cite{Bennewitz1964}. The
scattering operator as well as the rate operator are then diagonal in
the orientational DOFs since the Schr\"{o}dinger equation depends only
parametrically on the orientation \cite{Benjamin2015}.

Here, we are only interested in the c.m. dynamics of the molecule and,
thus, trace out the environmental as well as the orientational DOFs.
It follows from the monitoring approach
\cite{Hornberger2007,PhysRevA.82.042111,Hornberger2007} that the
motional state of a slowly rotating molecule in a gaseous environment
is described by a master equation
\begin{equation}
\partial_{t}\rho=\frac{1}{i\hbar}[\mathsf{H},\rho]+\mathcal{R}\rho+\mathcal{L}\rho,
\label{eq:master}
\end{equation}
where $\mathsf{H}={\sf P}^2/2M$ is the free Hamiltonian. The two superoperators in \eqref{eq:master}
can be given as
\begin{eqnarray}
\mathcal{R} \rho & = &i \left \langle
\mathrm{Tr}_{\mathrm{g}}\left(\left[\mathsf{\Gamma}^{1/2}\mathrm{Re}\left(\mathsf{T}\right)\mathsf{
\Gamma}^{1/2}, \rho\otimes\rho_{\mathrm{g}}\right]\right) \right \rangle,\label{eq:R_rho}\\
\mathcal{L} \rho & = & \Bigl \langle \mathrm{Tr}_{\mathrm{g}} \Bigl
(\mathsf{T}\mathsf{\Gamma}^{1/2}\rho\otimes\rho_{\mathrm{g}}\mathsf{\Gamma}^{1/2}\mathsf{T}^{\dagger
}  \nonumber \\
 &  &  -\frac{1}{2} \left
\{\rho\otimes\rho_{\mathrm{g}},\mathsf{\Gamma}^{1/2}\mathsf{T}^{\dagger}\mathsf{T}\mathsf{\Gamma}^{
1/2} \right \} \Bigr ) \Bigr \rangle,\label{eq:el_rho}
\end{eqnarray}
where ${\sf T} = {\sf T}(\Omega)$ is the non-trivial part of the
scattering operator,
$\mathsf{S}(\Omega)=\mathsf{1}+i\mathsf{T}(\Omega)$, and the
expectation value of the rate operator $\mathsf{\Gamma} =
\mathsf{\Gamma}(\Omega)$ gives the scattering rate. Here, the angular
brackets $\langle \cdot \rangle$ denote the orientational average with
respect to the rotation state of the molecule. We note that the
orientational coherences $\matel{\Omega}{\rho_{\rm M}}{\Omega'}$ do
not contribute to the master
equation~\eqref{eq:master} because the rate operator
and the scattering operator are diagonal in the orientational DOFs.

In most experimentally relevant cases the initial rotation state is thermal, implying that the
orientational DOFs are homogeneously distributed. Since the gas distribution $\mu(\vert {\bf p}
\vert)$ is approximately isotropic, it is natural to assume that the orientational distribution
remains homogeneous for all times, i.e.~the orientational diagonal elements of the total molecular
state $\rho_{\rm M}$ are $\matel{\Omega}{\rho_{\rm M}}{\Omega} = \rho / (8 \pi^2)$. Then the
orientational average of the function $K(\Omega)$ is
\begin{equation}
 \left \langle K(\Omega) \right \rangle = \int_{S} \frac{d\Omega}{8 \pi^2} K(\Omega),
\end{equation}
with $S = S_2 \times S_1$ being the configuration space of the orientational DOFs.

The fact that the molecular mass $M$ fairly exceeds the mass $m$ of a gas atom allows us to further
simplify Eq.~\eqref{eq:master}. In the limit that $m/M$ vanishes, Eq.~(\ref{eq:R_rho}) gives a
constant energy shift \cite{Vacchini2009}, which can be ignored,f and the second term
\eqref{eq:el_rho} describes decoherence in the position representation
\cite{Vacchini2009,Hornberger2008},
\begin{eqnarray}
\lim_{m/M\to0}\matel{\mathbf{R}_{1}}{\mathcal{L}\rho}{\mathbf{R}_{2}} & = & - \gamma \left [ 1 -
\eta(\mathbf{R}_{1}-\mathbf{R}_{2}) \right ] \notag \\
 && \times \matel{\mathbf{R}_{1}}{\rho}{\mathbf{R}_{2}}.\label{eq:stoss_deco}
\end{eqnarray}
This expression can be derived following the same steps \cite{Vacchini2009} as for a point particle.
The scattering rate $\gamma$ in \eqref{eq:stoss_deco} can be given as
\begin{equation}
\gamma = \frac{n_{\mathrm{g}}}{m} \int d^3\mathbf{p} \: p \mu( \vert \mathbf{p}+ m {\bf
v}_{\rm M} \vert) \left \langle \sigma_{\mathrm{tot}}(p) \right \rangle,\label{eq:gamma}
\end{equation}
where $p = \vert {\bf p} \vert$ is the length of the momentum vector and $n_{\mathrm{g}}$ is the
density of the gas. Equation \eqref{eq:stoss_deco} also involves the decoherence function
\begin{eqnarray}
\eta({\bf R}) & = & \frac{1}{N}\int_0^\infty dp \: p^{3}\int_{S_2}
d^2\mathbf{n}\int_{S_2}d^2\mathbf{n'}\mu( \vert \mathbf{p}+ m {\bf v}_{\rm M}
\vert)\: \notag \\
 && \times
\langle|f(p,\mathbf{n}\cdot\mathbf{n'})|^{2}\rangle\exp\left(\frac{ip}{\hbar}(\mathbf{n}
-\mathbf{n'})\cdot {\bf R} \right),\label{eq:eta}
\end{eqnarray}
where $\mathbf{n}$ and $\mathbf{n}'$ are the incoming and outgoing directions of the scattered gas
atom, respectively, and $S_2$ denotes the surface of the unit sphere. Here, the normalization
constant $N$ ensures that $\eta(0)=1$, so that the diagonal elements of $\rho$ are preserved by
Eq.~\eqref{eq:stoss_deco}. The fact that the differential scattering cross section $\langle \vert
f(p,{\bf n} \cdot {\bf n}' ) \vert^2 \rangle$ depends on the angle between incoming and outgoing
momentum rather than the individual directions is  due to the trace over the orientational DOFs
performed to derive Eq.~\eqref{eq:master}. The Fourier transform of the decoherence function
$\eta({\bf R})$ gives the probability distribution of transferred momentum of a single collision
\cite{Hornberger2004}.

Note that the thermal gas distribution in Eqs.~\eqref{eq:gamma} and
\eqref{eq:eta} is shifted by $m {\bf v}_{\mathrm{M}}$ due to the
molecule's velocity. Since the most probable momentum $p_{\rm g} =
\sqrt{2 m k_{\rm B} T}$ in a thermal gas distribution $\mu(p)$ is
usually much higher than the momentum $m v_{\rm M}$,
Eqs.~\eqref{eq:gamma} and \eqref{eq:eta} can be expanded in orders of
$m v_{\rm M} / p_{\rm g}$. This yields
\begin{equation}
 \gamma = \frac{4 \pi n_{\rm g}}{m} \int_0^\infty d p \: p^3 \mu(p) \left \langle
\sigma_{\rm tot}(p) \right \rangle \left [ 1  + \mathcal{O} \left ( \frac{m^2 v_{\rm M}^2}{p_{\rm
g}^2} \right ) \right ],
\end{equation}
and
\begin{eqnarray}
 \eta({\bf R}) & = & \frac{8 \pi^2}{N} \int_0^\infty dp\: p^3 \mu(p) \int_0^\pi d
\theta \: \sin \theta \langle|f(p, \cos \theta)|^{2}\rangle\, \notag \\
 && \times \mathrm{sinc}\left[\sin\left(\frac{\theta}{2}\right ) \frac{2p \vert {\bf R}
\vert}{\hbar}\right] \left [ 1 +\mathcal{O}\left ( \frac{m^2 v_{\rm M}^2}{p_{\rm g}^2} \right )
\right ]\label{eq:deco_function_cal1},
\end{eqnarray}
with $\cos \theta = {\bf n} \cdot {\bf n}'$. It is demonstrated in Appendix \ref{sub:linear_order}
that
the linear order vanishes.

In what follows, we will specify the scattering rate $\gamma$ and the decoherence function
$\eta({\bf R})$ for the case of a polar or anisotropically polarizable molecule scattering with
polarizable spherical particles.

\section{Cross Sections for polar molecule-atom scattering \label{sec:crosssec}}

It is the aim of this section to evaluate the total and the differential scattering cross sections
for the collision between a gas atom and a polar molecule. Although we are mainly interested in the
dipole-induced dipole interaction, we consider the more general homogeneous  potential
\begin{equation}
V \left (r,  \cos \Theta \right )=-\frac{C}{r^{s}} \left ( 1+a  \cos^2 \Theta  \right )
\label{eq:potential},
\end{equation}
which also includes the anisotropic van der Waals interaction
\cite{StoneBook}. Here, $\Theta$ is the angle between the molecule's
orientation ${\bf m}(\Omega)$ and the relative c.m. coordinate ${\bf
r}$, i.e. $\cos \Theta = {\bf m} \cdot {\bf r} / r$, and $r = \vert
{\bf r} \vert$ is the distance between the gas atom and the c.m. of the
molecule. The parameter $a \geq 0$ quantifies the anisotropy of the
interaction potential and $C$ is the interaction strength. For
example, the dipole-induced dipole interaction is described by $s = 6$
and $C = \alpha_0 d_0^2 / 32 \pi^2 \varepsilon_0^2$ with $a = 3$,
where $\alpha_0$ is the atomic polarizability and $d_0$ is the
molecular dipole moment. In this case ${\bf m}$ is the orientation of
the dipole moment.

The total and the differential scattering cross sections for a fixed molecular orientation ${\bf m}$
can be calculated in the eikonal approximation, also referred to as Schiff's approximation
\cite{Schiff1956}. This approximation provides reliable results for small angle scattering
\cite{Bennewitz1964}, i.e.~for soft collisions, which we will show to be most important for the
decoherence function. Choosing the initial relative momentum along the $z$-axis, ${\bf p} = p {\bf
e}_z$, the orientation dependent scattering amplitude can be expressed as \cite{Schiff1956}
\begin{widetext}
    \begin{equation}
  f(p, {\bf n}'; {\bf m}) = -\mathrm{i}\frac{p}{2\pi\hbar} \int_{\mathbb{R}^2} \mathrm{d}^2 {\bf b}
  \: e^{-\mathrm{i} p {\bf n}_\bot' \cdot {\bf b} / \hbar}  \left ( \exp \left [ - \frac{\mathrm{i}
  m}{\hbar p} \int_{-\infty}^\infty \mathrm{d} z\: V \left ( \sqrt{b^2 + z^2}, \frac{({\bf b} + z
{\bf  e}_z ) \cdot {\bf m}}{\sqrt{b^2 + z^2}} \right )  \right ] - 1\right ), \label{eq:f_amp}
  \end{equation}
  where ${\bf b}$ is the impact vector in the $xy$-plane and ${\bf n}'_\bot$ are the $xy$-components
  of the direction of outgoing momentum $p {\bf n}'$. Here, the scattering phase is calculated by
  integrating the interaction potential for a fixed molecular orientation ${\bf m}$ along the
straight
  trajectory of the gas particle.

  The total scattering cross section can be obtained by evaluating \eqref{eq:f_amp} in the forward
  direction, ${\bf n}' = {\bf e}_z$, according to the optical theorem. The orientationally averaged
  total cross section is then given by \cite{Bennewitz1964}
  \begin{eqnarray}
  \left \langle \sigma_{\mathrm{tot}}(p) \right \rangle & = & \sigma_{0}(p) \int_{S_2}
  \frac{d^2 {\bf m}}{4 \pi} \: \left [1+\frac{a}{s} \left ( {\bf m} \cdot {\bf e}_z \right
  )^2  +\frac{a(s-1)}{s} \left ( {\bf m} \cdot {\bf e}_x\right )^2
  \right]^{2/(s-1)},\label{eq:sig_tot}
  \end{eqnarray}
  where
  \begin{equation}
  \sigma_{0}(p)=2\pi\sin\left(\frac{\pi}{2}\frac{s-3}{s-1}\right)\Gamma\left(\frac{s-3}{s-1}
  \right)\left(\frac{\sqrt{\pi} m C}{\hbar p}\frac{\Gamma[(s-1)/2]}{\Gamma(s/2)}\right)^{2/(s-1)},
  \end{equation}
\end{widetext}
is the cross section resulting from the isotropic part of the interaction potential, i.e. for $a =
0$ in Eq.~\eqref{eq:potential}  \cite{LandauBook}.
The orientational average in Eq.~\eqref{eq:sig_tot} can be well approximated (see
Appendix \ref{app:diff_scatt}), and one obtains the compact expression
\begin{equation}
\langle\sigma_{\mathrm{tot}}(p)\rangle=\sigma_{0}(p)\left(1+\frac{a}{3}\right)^{2/(s-1)}.
\label{eq:sig_tot2}
\end{equation}
Thus the anisotropy of the potential \eqref{eq:potential} enhances the total cross section by a
constant factor.

We now turn our attention to the differential scattering cross section. In general, the scattering
amplitude \eqref{eq:f_amp} cannot be evaluated in closed form. However, for small angle scattering
events \cite{PaulyDiffSig}, for which ${\bf n}' \simeq {\bf n}$, the scattering amplitude can be
evaluated by expanding the plane wave in Eq.~\eqref{eq:f_amp} up to second order in ${\bf n}_\bot
\cdot {\bf b}$. A straightforward derivation, presented in Appendix \ref{app:diff_scatt}, gives the
orientation averaged differential scattering cross section
\begin{equation}
 \langle|f(p, \cos \theta)|^{2}\rangle = A(p) \left  [ 1 - \left (
\frac{\theta}{\theta_{*}(p)}\right)^{2} + \mathcal{O}(\theta^4)  \right ] \label{eq:dif_sig}
\end{equation}
where, making the same approximations as above, we abbreviated
\begin{eqnarray}
A(p) & = & \left(\frac{p\langle\sigma_{\mathrm{tot}}(p)\rangle}{4\pi\hbar \cos [\pi  /
(s-1)]}\right)^{2}, \hspace{3cm} \label{eq:A} \\
\theta_{*}(p) & = & \frac{\hbar}{p} \sqrt{\frac{8\pi}{\langle\sigma_{\mathrm{tot}}
(p)\rangle}}\Gamma\left(\frac{s-3}{s-1}\right)\Gamma\left(\frac{s-5}{s-1}\right)^{-1/2}.
\label{eq:thetas}
\end{eqnarray}

Thus, the differential cross section as a function of $\theta$ decays quadratically in the forward
direction. However, its functional dependence for all $\theta\in[0,\pi]$ is required for the
calculation of the decoherence function (\ref{eq:eta}). One possibility to overcome this is to
approximate the cross section by a Gaussian curve with amplitude $A$ (\ref{eq:A}) and width
$\theta_{*}$ (\ref{eq:thetas}) \cite{PaulyDiffSig}
\begin{equation}
\langle|f(p,\cos \theta)|^{2}\rangle\approx A(p) \exp\left
[-\left(\frac{\theta}{\theta_{*}(p)}\right)^{2}\right].\label{eq:dif_sig_gaus}
\end{equation}
This is certainly a crude approximation, and even though it was demonstrated \cite{PaulyDiffSig}
that Eq.~(\ref{eq:dif_sig_gaus}) gives reliable results for soft scattering, $\theta\ll1$, it does
not contain the correct asymptotic behavior $\langle |f|^2 \rangle \propto \theta^{-2(s+1)/s}$ for
hard scattering $\theta \lesssim 1$ \cite{LandauBook,Mount1973}. However, the important properties
of the decoherence function, i.e.,its width and its asymptotic behavior as $\vert {\bf R} \vert \to
\infty$, are mainly determined by soft collisions, since hard collisions cause almost complete
decoherence and are therefore mainly captured by the scattering rate. Importantly, this scattering
rate is determined without the small-angle approximation; see Eq.~\eqref{eq:sig_tot}. We remark that
the differential cross section can also show oscillating behavior (glory and rainbow effect)
\cite{Berry1972} and various resonances \cite{Drake2005}, but neglecting these is justified, since
they average out when evaluating the 
decoherence function (\ref{eq:eta}).

Having discussed the validity of the approximations used, we are now in the position to compute the
decoherence function \eqref{eq:eta} and the scattering rate \eqref{eq:gamma} in the next section.

\section{Decoherence function and scattering rate} \label{sec:decscatt}

In order to calculate the scattering rate, we insert the total scattering cross section
\eqref{eq:sig_tot2} into Eq.~\eqref{eq:gamma} and obtain for the leading order contribution in $m
v_{\rm M}/p_{\rm g}$
\begin{equation}
\gamma = \frac{2 n_{\mathrm{g}}p_{\rm g}}{m\sqrt{\pi}} \Gamma\left(\frac{2s - 3}{s-1}\right)
\langle\sigma_{\mathrm{tot}}(p_{\rm g})\rangle.
\end{equation}
As expected, the scattering rate is proportional to $n_{\rm g} p_{\rm g} \sigma_{\rm tot} / m$,
which is the effective current of gas atoms passing through the total scattering cross section. For
the case of the dipole-induced dipole interaction, $s = 6$, the rate is given by 
\begin{equation}
 \gamma_{\rm d-id} = \frac{2^{7/5}}{\sqrt{\pi}} \Gamma\left(\frac{9}{5}\right) \frac{n_{\mathrm{g}}
p_{\rm g} \sigma_0(p_{\rm g})}{m}.
\end{equation}
In particular, it is proportional to  $n_{\rm g} \alpha_0^{2 / 5} d_0^{4 / 5} (T / m)^{3/10}$ and
thus depends only weakly on the gas temperature $T$. In comparison to the scattering rate of a
spherical molecule \cite{Hornberger2004}, one observes an additional prefactor of $2^{2/5} \simeq
1.3$ due to the anisotropic contribution to the interaction potential \eqref{eq:potential}.

The evaluation of the decoherence function $\eta({\bf R})$ is slightly more complicated. In
accordance with the small angle approximation utilized in the calculation of the differential
scattering cross section \eqref{eq:dif_sig_gaus}, we replace in Eq.~\eqref{eq:eta} $\sin \theta$ by
$\theta$ and extend the integration boundary to infinity. Thus, the leading order contribution in $m
v_{\rm M}/p_{\rm g}$ is
\begin{eqnarray} \label{eq:eta2}
\eta({\bf R}) & = & \int_{0}^{\infty}d\xi\: \nu(\xi) D\left[\frac{p_{\rm g}\vert {\bf R}
\vert \theta_{*}(p_{\rm g})}{2\hbar}\xi\right],
\end{eqnarray}
where $\xi = (p / p_{\rm g})^{1/(s - 1)}$ is a dimensionless integration variable, and we defined
the probability distribution $\nu(\xi) = 2(s-1) \xi^{4 s -7} \exp \left ( - \xi^{2 ( s- 1)}  \right
) / \Gamma [ (2 s - 3) / (s -1) ]$ as well as the auxiliary function
\begin{equation}
D(x)= \frac{e^{-x^2}}{x} \int_0^x d\xi\: e^{\xi^2}.
\end{equation}
The function $D(x)$ is symmetric and monotonically decreasing, with its maximum value $D(x = 0)=1$,
so that the decoherence function fulfills $\eta(0) = 1$. For large arguments it decreases as $D(x)
\sim 1/x^{2}$. We remark that it is related to Dawson's integral $F(x)$ by $D(x) = F(x)/x$
\cite{Abramowitz}.

The function $\nu(\xi)$ is sharply peaked near its mean $\xi_s =
\Gamma [ 2  - 1/2(s - 1)]/\Gamma [ 2  - 1/(s - 1)]$, so that the
function $D(x)$ is approximately linear in this region. This allows us
to approximate the integral \eqref{eq:eta2} to obtain the decoherence
function
\begin{equation} \label{eq:eta_super}
\eta({\bf R}) \simeq D\left[\frac{\xi_s p_{\rm g}  \theta_{*}(p_{\rm g}) \vert {\bf R}
\vert}{2\hbar}\right].
\end{equation}

The decoherence function describes the decay of the coherences by a single scattering event
\cite{Hornberger2004}. Its characteristic width is given by
\begin{equation}
 w_\eta = \frac{2 \hbar}{\xi_s p_{\rm g} \theta_*(p_{\rm g})} \propto \sqrt{\langle \sigma_{\rm
tot}(p_{\rm g}) \rangle} \propto  p_{\rm g}^{-1/(s - 1)},
\end{equation}
i.e.~it decreases with increasing momentum $p_{\rm g}$, where decoherence is more pronounced. The
decoherence function \eqref{eq:eta} shows the asymptotic behavior $\eta({\bf R}) \to 0$ for $\vert
{\bf R} \vert \gg w_\eta$, i.e. the coherence of distant spatial superpositions gets fully
destroyed.

The Fourier transform $\widetilde{\eta}({\bf P})$ of the decoherence function gives the distribution
of momentum kicks \cite{Hornberger2004}. In the present case, it can be evaluated explicitly as
\begin{eqnarray}
 \widetilde{\eta}({\bf P}) & = & \frac{1}{(2 \pi \hbar)^3}\int_{\mathbb{R}^3} d^3{\bf R} \:
\eta({\bf R}) e^{- i {\bf R} \cdot {\bf P} / \hbar} \notag \\
  & = & \frac{1}{2 \pi\vert {\bf P} \vert} \left( \frac{w_\eta}{2 \hbar} \right )^2  \exp \left [ -
\left ( \frac{\vert {\bf P} \vert w_\eta}{2 \hbar} \right )^2 \right ], \label{eq:eta_fourier}
\end{eqnarray}
which involves the most probable transferred momentum $\sqrt{2} \hbar / w_\eta$. Note that the
singularity of $\widetilde{\eta}({\bf P})$ at ${\bf P} = 0$ reflects the fact that the decoherence
function \eqref{eq:eta_super} is not normalizable. Nevertheless, its Fourier transform
\eqref{eq:eta_fourier} is normalized in accordance with $\eta(0) = 1$.

For example, in the case of the dipole-induced dipole interaction, the
width of the decoherence function $w_\eta \approx 0.6 \sqrt{\langle
\sigma_{\rm tot}(p_{\rm g}) \rangle}$ is typically in the of range
nanometers. In the case of helium atoms, $\alpha_0/4 \pi \varepsilon_0
= 0.2 ~$\AA${}^3$, moving with $p_{\rm g}/m = 10^3$~m/s, and the
molecular dipole moment $d_0 = 5$ Debye, one obtains $w_\eta \simeq
0.5$ nm.

\section{Application to far-field matter-wave interferometry\label{sec:application}}

\begin{figure}
\includegraphics[scale=0.35]{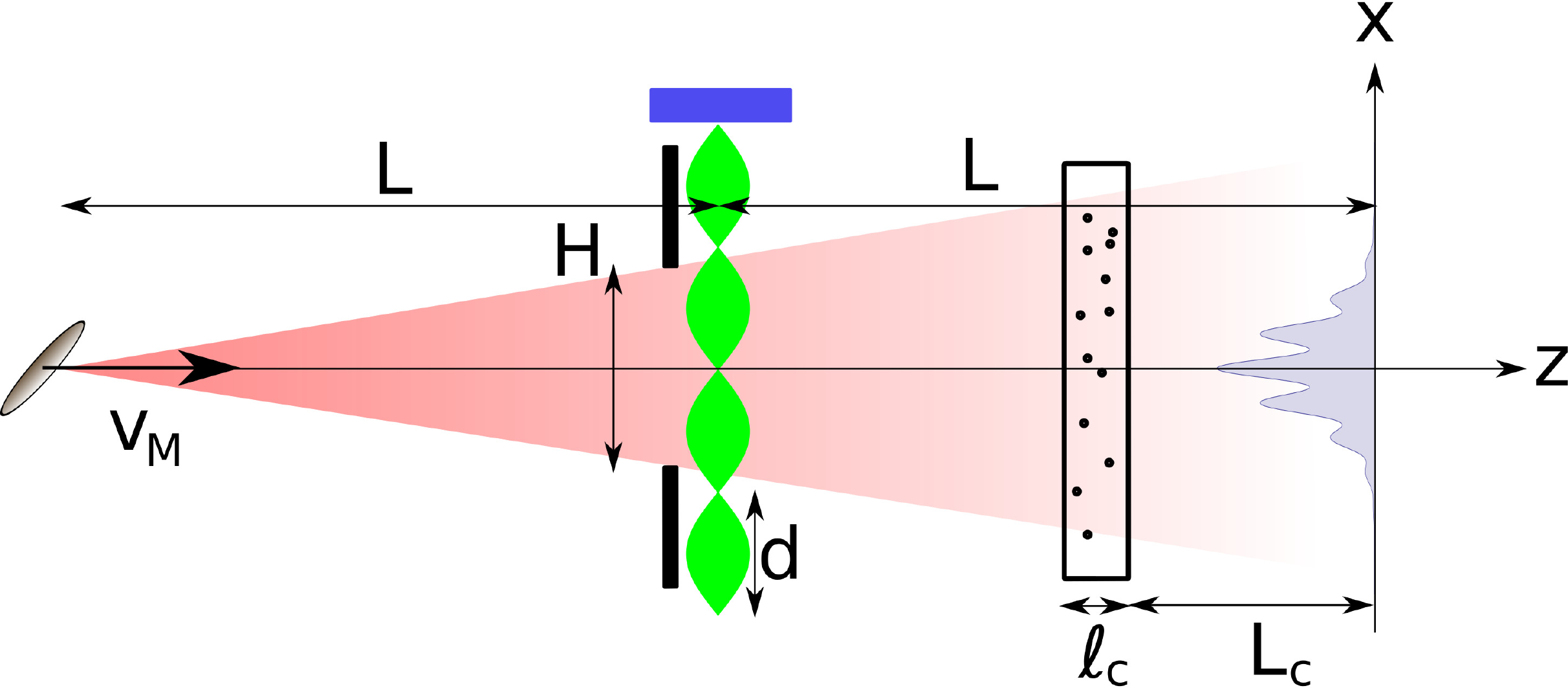}\caption{Schematic of a far-field interference
experiment with a dedicated collision chamber. The molecules are emitted with longitudinal velocity
$v_M$ from a point-like source ($z = 0$), propagate the distance $L$ to the grating ($z = L$), and
are detected on the screen ($z = 2L$) after traversing the collision chamber of width $\ell_{\rm c}$
placed at the distance $L_{\rm c}$ in front of the screen.} \label{fig:schematic}
\end{figure}

In order to illustrate the results of the previous sections, we consider a far-field matter-wave
experiment with massive molecules. The particles are emitted with longitudinal velocity ${\bf
v}_{\rm M} = v_{\rm M} {\bf e}_z$ from a point-like source, propagate freely over the distance $L$
to the diffraction grating with period $d$, and then propagate again for the distance $L$ to the
detection screen \cite{Arndt1999}, see Fig.~\ref{fig:schematic}. Since the longitudinal kinetic
energy exceeds the average interaction strength as well as the transverse kinetic energy, the
interaction with the grating can be described in the eikonal approximation and it suffices to
consider the transverse state \cite{Nimmrichter2008}.

We choose our coordinate system such that $x$ denotes the grating axis and $z$ is the flight
direction. The $y$ dependence can be neglected in most matter-wave experiments since the extension
of the grating in this direction is much larger than the spatial coherence of the particle. In the
case of a point source, the interference pattern at the screen can be given as \cite{BornWolfBook}
\begin{equation}
w_{0}(x) \propto \left \vert\int_{-\infty}^\infty dx'\exp\left [-i\frac{2\pi x' (
x - x')}{d \Delta x} \right ] t(x') \right \vert^{2},
\end{equation}
up to an irrelevant prefactor. Here $\Delta x = 2\pi\hbar L/dMv_{{\rm M}}$ is the separation of
diffraction peaks in the far field, $H^2 / d \Delta x \ll 1$, and we defined the grating
transformation $t(x)$, which describes the passage through the grating in the eikonal approximation,
$\rho \to \hat t \rho \hat t^\dagger$. In what follows, we consider a far-detuned laser grating of
width $H$, $t(x)=\exp [i\phi_{0}\cos^{2}(\pi x/d) ] \Theta(H/2-|x|)$, which acts as a pure
phase grating with the maximal phase shift $\phi_{0}$. In the far field, $H^2 \ll d \Delta x$, the
interference pattern is given by the modulus squared of the Fourier transform of the grating
function $t(x)$. For the sake of a clear presentation, we neglect the influence of photon absorption
\cite{Cotter2015}.

For a typical width of the decoherence function \eqref{eq:eta_super},
$w_\eta \simeq 0.5$ nm, the most probable transferred momentum
$\sqrt{2} \hbar / w_\eta$ according Eq.~\eqref{eq:eta_fourier} is much
larger than the grating momentum $2 \pi \hbar /d$. For instance, for
$d = 200$ nm, the ratio is $d / \sqrt{2} \pi w_\eta \simeq 90$. This
implies that collisions occurring far away from the detection screen
reduce the signal because a fraction of the molecules are effectively
kicked out of the finite detection range while the shape and contrast
of the interference pattern are almost not affected. Hence, the effect
of collisional decoherence on the signal is best investigated if the
collisions take place close to the screen.

Therefore, we consider a collision chamber of width $\ell_{\rm c}$
placed at the distance $L_{\rm c}$ in front of the detector, see
Fig.~\ref{fig:schematic}. We take this chamber to be filled with a
mono-atomic gas of density $n_{\rm g}$ and temperature $T$, and the
interaction between the polar molecule and the gas particles to be
described by the dipole-induced dipole interaction. The range of the
homogeneous potential \eqref{eq:potential} can be estimated to be
$R_{\mathrm{w}}\simeq (2 m C/\hbar^{2})^{1/(s-2)}$, where $m$ is the
mass of the gas particle. For example, for helium atoms, $m = 4$ amu,
at room temperature, $p_{\rm g} / m \simeq 10^3$\,m/s, and a molecule
with dipole moment $d_0 = 5$ D, the characteristic time scale of the
scattering process is approximately $\tau_{\rm c} = m R_{\rm w}/p_{\rm
g}\simeq 0.7$ ps. On the other hand, the rotational period of a linear
rigid molecule of mass $M=10^{3}$ amu and length $3$ nm which has the
internal temperature $T_{\mathrm{M}}=1000$ K, is approximately
$\tau_{\mathrm{rot}}\simeq 40$ ps, and thus $\tau_{\rm c} / \tau_{\rm
rot} \simeq 50$. Hence, the molecule is almost static during the
interaction and the sudden approximation is well justified.

It is shown in Appendix \ref{sec:Molecular-density-distribution} that the
interference pattern can be calculated by
\begin{equation} \label{eq:signal}
w(x)=e^{-\gamma\ell_{\rm c}/v_{{\rm M}}}\left[w_{0}(x)+ \int_{-\infty}^\infty d x'
w_{0}(x') h(x - x' )\right],
\end{equation}
where we defined
\begin{eqnarray}
h(x)  &=&  \int_{-\infty}^\infty \frac{dq}{2\pi\hbar}\:e^{-iqx/\hbar} \hspace{4cm}
\notag \\
 & \times& \left \{\exp\left [ \frac{\gamma}{v_{\mathrm{M}}}\int_{0}^{ \ell_{{\rm
c}}} dz\:\eta\left(\frac{q (L_{{\rm c}} + z)}{Mv_{{\rm M}}}\right )\right ] - 1 \right \}.
\end{eqnarray}
We note that the relation \eqref{eq:signal} preserves the normalization of $w(x)$. The first term on
the right-hand side of Eq.~\eqref{eq:signal} describes the local signal loss, while the second term
describes the modification of the shape of the interference pattern due to collisions. If the
collision chamber is far away from the detection screen, $L_{\rm c} \gg L w_\eta/ d$, the first term
dominates the interference signal and the intensity is locally reduced but the shape of the fringes
remains unchanged.

\begin{figure}
\includegraphics[scale=0.4]{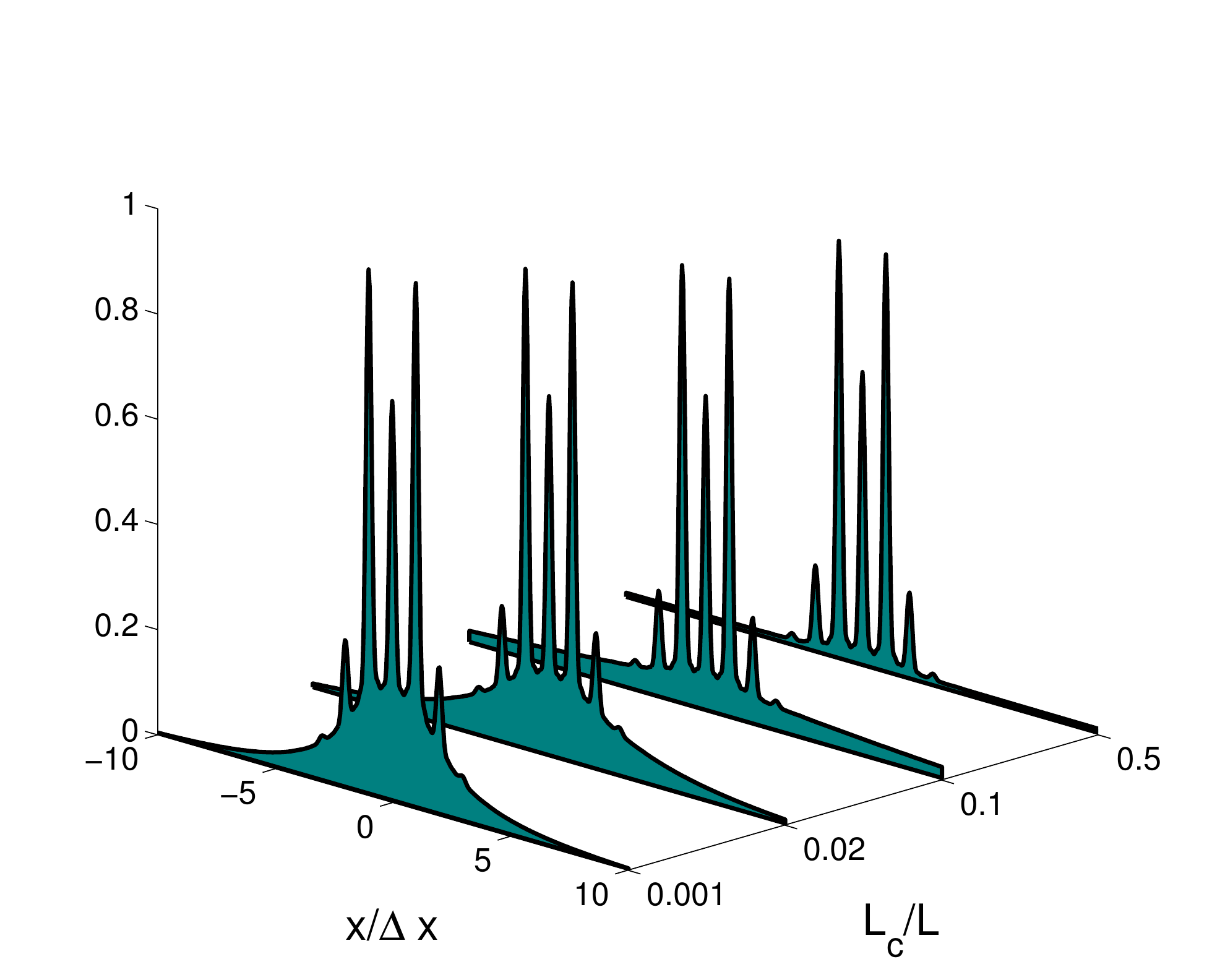}\caption{Far-field
interference pattern of a polar molecule in the presence of a collision chamber of width $\ell_{\rm
c} /L = 0.05$, which is placed in front of the detector
at the distances $L_{{\rm c}}/L=0.001$, $0.02$, $0.1$ and $0.5$.
The molecules of mass $840$ amu move with velocity $v_{{\rm M}}=50$
m/s,  and have a dipole moment of $5$ Debye. The collision
chamber is filled with Helium gas at $300$ K and
$5$ mPa. The atomic polarizability of Helium atoms is taken to be \cite{Handbook}
$\alpha/4\pi\varepsilon_{0}=0.2 \:$\AA${}^{3}$.
The phase shift and the width of the diffraction grating are chosen
to be $\phi_{0}=\pi$ and $H/d=5$.}\label{fig:Far_deco_fig}
\end{figure}

In Fig.~\ref{fig:Far_deco_fig} we show the influence of the collision
chamber on the far-field interference pattern of a polar molecule
diffracted from a pure phase grating as a function of the distance
$L_{\rm c}$ between the detection screen and the chamber. The
collisions with gas atoms induce a position-dependent underground. The
width $\Delta I$ of this underground can be estimated assuming that
the collisions occur exactly in the middle of the collision chamber,
\begin{equation}
\frac{\Delta I}{\Delta x} \simeq \frac{d}{\sqrt{2}\pi w_\eta} \frac{L_{{\rm c}}+\ell_{\rm c}/2}{L}.
\end{equation}
This relation reflects the fact that in order to investigate the effect of decoherence, the distance
between the collision chamber and the detector must be of the order $L_{{\rm c}}/L\simeq w_\eta /d$.

The fact that the exponential reduction of the signal visibility is proportional to the scattering
rate $\gamma$ allows measuring the relative interaction strengths $C^{(1)}/C^{(2)}$ by performing
the experiment for two different molecules. For example, if the dominant interaction for both
molecules is the dipole-induced dipole interaction, the ratio of the scattering rates obtained for
the molecules is $\gamma_{\rm d-id}^{(1)} / \gamma_{\rm d-id}^{(2)} = (d_0^{(1)} / d_0^{(2)} )^{4 /
5}$. This can be used as a measurement of molecular dipole moments in the gas phase.

\section{Conclusion} \label{sec:conc}

We derived the master equation for the center-of-mass collisional decoherence of a slowly rotating
polar molecule in a thermal gas. Based on the orientationally averaged differential and total
scattering cross section of a single collision, we calculated the decoherence function and the
scattering rate for the case that the orientation is homogeneously distributed. The decoherence
function describes the decay of the coherences and provides the most probable transferred momentum.
In the case of the dipole-induced dipole interaction, its width is of the same order of magnitude as
the square root of the total scattering cross section.

As an application, we considered a far-field matter-wave experiment
with an additional collision chamber placed between the grating and
the detector and filled with a mono-atomic gas. If the chamber is far
away from the detection screen, the interference intensity decreases
exponentially with the number of scattering events. Comparing the
scattering rate of two different molecules may then allow one to
measure the relative strength of molecule-gas interaction. In the case
that both molecules are polar, one can thus extract their relative
dipole moments. On the other hand, if the collision chamber is placed
closer to the detection screen, one starts to observe the
modification of the shape of the interference fringes due to
collisional decoherence.

\begin{acknowledgments}
 We acknowledge support from the European Commission within NANOQUESTFIT (Contract No. 304886).
\end{acknowledgments}

\begin{appendix}

\section{Influence of the anisotropy of the thermal gas distribution} \label{sub:linear_order}

We consider here the angular integration required to evaluate the decoherence function
\eqref{eq:eta}, i.e.,
\begin{eqnarray} \label{eq:I}
I & = & \int_{S_2}d^2\mathbf{n}
\int_{S_2} d^2\mathbf{n'}\:\mu(\vert
\mathbf{p}+\mathbf{p}_{0}
\vert)\langle|f(p,\mathbf{n}\cdot\mathbf{n'})|^{2}\rangle \notag \\
 && \times \exp\left [i\zeta \mathbf{u} \cdot (\mathbf{n}-\mathbf{n'})\right ], 
\end{eqnarray}
with the unit vector $\mathbf{u} = (\mathbf{R}_{2}-\mathbf{R}_{1})/ \vert
\mathbf{R}_{2}-\mathbf{R}_{1} \vert$ and the constant $\zeta = \vert \mathbf{R}_{2}-\mathbf{R}_{1}
\vert p / \hbar$. If the momentum shift $\vert {\bf p}_0 \vert = m v_{\rm M}$ is small in comparison
to the most probable momentum $p_{\rm g}$ in the thermal distribution $\mu(\vert {\bf p} \vert)$,
one can expand 
\begin{eqnarray}
\mu( \vert {\bf p} + {\bf p}_0 \vert) &  = & \mu(p) + ({\bf p}_0 \cdot {\bf n}) \partial_p \mu(p)
\notag \\
 && \times \left [ 1 + \mathcal{O} \left ( \frac{m v_{\rm M}}{p_{\rm g}} \right ) \right ],
\end{eqnarray}
where we used that ${\bf p} = p {\bf n}$. The zero-th order contribution is spherically symmetric,
and the integration result is independent of the direction ${\bf u}$, i.e.,
\begin{eqnarray}
 I_0 & = & 8 \pi^2 \mu(p) \int_0^\pi d\theta\: \sin \theta \langle|f(p,\cos
\theta)|^{2}\rangle \notag \\
  && \times \mathrm{sinc} \left [2 \zeta \sin \left ( \frac{\theta}{2} \right ) \right ], \, 
\end{eqnarray}
where $\cos \theta = {\bf n} \cdot {\bf n'}$. The first-order contribution is given by
\begin{eqnarray}
I_1 & = & [\partial_p \mu(p)] {\bf p}_0 \cdot \int_{S_2} d^2\mathbf{n}\int_{S_2}
d^2\mathbf{n'}\:{\bf n}\langle|f(p,\mathbf{n}\cdot\mathbf{n'})|^{2}\rangle \notag \\
 && \times \exp\left [i\zeta {\bf u} \cdot (\mathbf{n}-\mathbf{n'})\right ].
\end{eqnarray}
Simultaneously substituting ${\bf n} \to -{\bf n}$ and ${\bf n}' \to -{\bf n}'$ yields that the real
part of $I_1$ is equal to zero. On the other hand, the imaginary part of $I_1$ can be expressed as
\begin{equation}
 \frac{I_1 - I_1^*}{2 i} \propto {\bf p}_0 \cdot \nabla_{\bf u} I_0 = 0,
\end{equation}
since $I_0$ is independent of the direction ${\bf u}$. Thus we demonstrated that the linear
contribution vanishes and the integral \eqref{eq:I} can be written as
\begin{equation}
 I = I_0 \left [ 1 + \mathcal{O} \left ( \frac{m^2 v_{\rm M}^2}{p_{\rm g}^2} \right ) \right ]
\end{equation}

\section{The small angle differential scattering cross section} \label{app:diff_scatt}

Here, we derive the small angle differential cross section from Eq.~\eqref{eq:f_amp} by expanding
the plane wave up to second order in ${\bf n}_\bot \cdot {\bf b}$ \cite{ChildBook}. For this purpose
it is beneficial to use spherical coordinates: The outgoing momentum points into the direction ${\bf
n}' = {\bf n}'(\theta,\phi)$, where $\theta$ and $\phi$ denote the polar and azimuthal angles and,
similarly, ${\bf m} = {\bf m}(\beta,\alpha)$. In addition, $\varphi$ is the angle between ${\bf b}$
and the $x$ axis.

In what follows, we will employ the following two approximations:
\begin{equation}
\frac{1}{2\pi}\int_{0}^{2\pi}d\phi(1+c\cos^{2}\phi)^{\mu} \approx
\left(1+\frac{c}{2}\right)^{\mu},\label{eq:app1-1}
\end{equation}
where $0 \leq c \lesssim 3$ and $0 \leq \mu \leq 1$, and
\begin{equation}
\frac{1}{2}\int_{0}^{\pi}d\phi\sin\phi(1+c\cos^{2}\phi)^{\mu} \approx
\left(1+\frac{c}{3}\right)^{\mu},\label{eq:app2-1}
\end{equation}
where $\vert c \vert < 1$ and $0 \leq \mu \leq 3$. The relative error does not exceed $3 ~\%$ for
the parameter range given above.

For the discussion to follow, it is important to note that the $z$ integration in
Eq.~\eqref{eq:f_amp} can be carried out explicitly to give
\begin{eqnarray}
&&\int_{-\infty}^\infty d z\: V \left ( \sqrt{b^2 + z^2}, \frac{({\bf b} + z {\bf e}_z )
\cdot {\bf m}}{\sqrt{b^2 + z^2}} \right ) \notag \\
&&= -\frac{\sqrt{\pi} C}{b^{s - 1}} \frac{\Gamma
\left[( s -
1) / 2 \right ]}{\Gamma(s/2)} \left [ G(\beta, \varphi - \alpha) \right ]^{(s-1)/2},
\end{eqnarray}
where
\begin{eqnarray}
 G(\beta, \varphi)  = \left( 1 + \frac{a}{s} \cos^2 \beta + \frac{a (s - 1)}{s} \cos^2 \varphi
\sin^2 \beta \right )^{2/(s - 1)}. \notag \\
\end{eqnarray}

Then, the zeroth order contribution $f_0(p,{\bf n'}; {\bf m})$ can be evaluated with the help of the
integrals \cite{Gradshteyn}
\begin{eqnarray}
&&\int_{0}^{\infty}db\:b^{m}\sin^{2}\left(\frac{A}{2b^{s-1}}\right) = \frac{A^{(m+1)/(s-1)}}{2(m+1)}
\notag \\
&& \hspace{1cm} \times\Gamma\left(\frac{s-m-2}{s-1}\right)
\sin\left(\frac{\pi}{2}\frac{s-m-2}{s-1}\right) \\
&&\int_{0}^{\infty}db\: b^{m}\sin\left(\frac{A}{b^{s-1}}\right) = \frac{A^{(m+1)/(s-1)}}{(m+1)}
\notag \\
&& \hspace{1cm} \times \Gamma\left(\frac{s-m-2}{s-1}\right)
\cos\left(\frac{\pi}{2}\frac{s-m-2}{s-1}\right),
\end{eqnarray}
for $A , m \in \mathbb{R}$, and $1 < (m + s)/(s - 1) < 3$ as well as $0 < (m + s)/(s - 1) <2$,
respectively. Thus, one obtains
\begin{eqnarray}
 f_0(p,{\bf n}'; {\bf m}) & = & \frac{p}{4 \pi \hbar} \exp \left (\frac{i \pi}{2} \frac{s - 3}{s-1}
\right ) \Gamma \left ( \frac{s - 3}{s-1} \right )  \notag \\
&\times& \left ( \frac{m \sqrt{\pi} C }{\hbar p}
\frac{\Gamma[(s -1)/2]}{\Gamma(s/2)} \right )^{2/(s-1)} \notag \\
&\times&  \int_0^{2 \pi} d \varphi \:G(\beta, \varphi ) .
\end{eqnarray}

In a similar fashion, the linear contribution vanishes due to the symmetry of the integrand. The
leading order correction is thus the quadratic term, which can be evaluated analogously to the above
calculation,
\begin{eqnarray}
 & & f_2(p,{\bf n}'; {\bf m}) \hspace{5cm} \notag \\
& & = - \frac{p}{16 \pi \hbar} \exp \left (\frac{i \pi}{2} \frac{s-
5}{s-1} \right ) \Gamma \left ( \frac{s - 5}{s-1} \right ) \notag \\ 
& &\times \left ( \frac{p \theta}{\hbar}\right )^2 \left ( \frac{m \sqrt{\pi} C }{\hbar p}
\frac{\Gamma[(s -1)/2]}{\Gamma(s/2)} \right
)^{4/(s-1)}\notag \\
 & & \times \left [ \sin^2(\phi - \alpha) \int_0^{2 \pi} d\varphi \: G^2(\beta,\varphi) +
\cos ( 2 \phi - 2\alpha) \right. \notag \\
& & \times \left. \int_0^{2 \pi} d\varphi \: \cos^2 \varphi G^2(\beta, \varphi)\right ].
\end{eqnarray}
Adding the two contributions and orientationally averaging the squared absolute value finally gives
Eq.~\eqref{eq:dif_sig}, where the constants $A$ and $\theta_*$ are given by
\begin{eqnarray}
A & = & \left(\frac{p \sigma_{\mathrm{0}}(p)}{4\pi
\hbar}\right)^{2}\left[1+\tan^{2}\left(\frac{\pi}{s-1}\right)\right] h_1(a),\label{eq:A_ave}\\
\theta_{*} & = & \frac{\hbar}{p}
\sqrt{\frac{8\pi}{\sigma_{\mathrm{0}}(p)}}\Gamma\left(\frac{s-3}{s-1}
\right)\left[\Gamma\left(\frac{s-5}{s-1}\right) h_2(a)\right]^{-1/2}.
\nonumber \\
 && \label{eq:ths_ave}
\end{eqnarray}
Here, we defined the two functions
\begin{equation}
h_1(a)=\frac{1}{8 \pi^2}\int_{0}^{\pi}d\beta\:\sin\beta \left  [ \int_0^{2 \pi} d
\varphi G(\beta,\varphi) \right ]^2,\label{eq:ABintegral-2}
\end{equation}
and
\begin{eqnarray}
h_2(a) & = & \frac{1}{4 \pi^2h_1(a) } \int_{0}^{\pi} d\beta \int_0^{2 \pi} d
\varphi d \varphi' \:\sin\beta\; \notag \\
 && \times G(\beta,\varphi) G^2(\beta,\varphi').\label{eq:ABintegral-1}
\end{eqnarray}

With the approximations presented above, the two functions \eqref{eq:ABintegral-2} and
\eqref{eq:ABintegral-1} can be written as
\begin{eqnarray}
 h_1(a) & = & \left ( 1 + \frac{a}{3} \right )^{4/(s-1)}, \\
 h_2(a) & = & \left ( 1 + \frac{a}{3} \right )^{2/(s-1)},
\end{eqnarray}
which finally gives Eqs. \eqref{eq:A} and \eqref{eq:thetas}, respectively.

\section{Molecular density distribution in far-field
interferometry\label{sec:Molecular-density-distribution}}

Here, we present the theoretical description of the molecular c.m.
dynamics in a far-field matter-wave experiment with a collision
chamber. A typical far-field setup consists of a molecular source ($z
= 0$), a diffraction grating ($z = L_1$) and a screen ($z = L_1 +
L_2$). Since the motional c.m. state of the molecule is separable in
$x$, $y$ and $z$ directions at all times \cite{Hornberger2004}, one
can reduce the problem to a single dimension and it suffices to treat
the dynamics in the $x$-direction. The motion in the $z$ direction is then
approximated by replacing $z = v_{\rm M} t$.

The interference pattern can be conveniently expressed with the help of the Wigner function $w(x,p)$
or, equivalently, with its characteristic function \cite{Bateman2014}
\begin{equation}
\chi(s,q)=\int_{-\infty}^\infty{\rm d}x \int_{-\infty}^\infty{\rm d}p\;
w(x,p)e^{i(qx-ps)/\hbar}.
\end{equation}
In particular, the solution of the master equation for collisional decoherence,
\begin{eqnarray}
\partial_t\rho(x,x') &= & \frac{1}{i
\hbar}\matel{x}{[\mathsf{H},\rho]}{x'} \notag
\\
 && +\gamma(t)[\eta(x-x')-1 ]\rho(x,x'),
\end{eqnarray}
with the free Hamiltonian $\mathsf{H}$ and the time-dependent scattering rate $\gamma(t)$ can be
given in terms of the characteristic function as
\begin{eqnarray}
&&\chi_{t}(s,q)=\chi_{0}\left(s-\frac{qt}{M},q\right) \hspace{3cm} \notag \\
&& \times \exp\left\{
\int_{0}^{t}d\tau\:\gamma(\tau)\left[\eta\left(s-\frac{q}{M}(t-\tau)\right)-1\right]\right\}.\label{
eq:propagation}
\end{eqnarray}
The state after time $t$, $\chi_{t}$, is described by multiplying the initial state $\chi_{0}$ by an
exponential reduction factor due to decoherence followed by the shearing transformation $s\to
s-qt/M$. The time dependence of the collision rate $\gamma$ may account for a spatially varying gas
density, $n_{\rm g}(z = v_{\rm M} t)$.

If the scattering occurs in a chamber placed between the diffraction grating and the detector, the
state in front of the detector reads
\begin{eqnarray}
\chi_{t_{1}+t_{2}}(s,q)&=&\chi'_{t_{1}}\left(s-\frac{q}{M}t_{2},q\right) \notag \\
&\times& \exp\Bigg\{\int_{0}^{t_{2}}d\tau\:\gamma(\tau+t_{1})  \notag \\
&\times&  \left[\eta\left(s-\frac{q}{M}(t_{2} -\tau)\right)-1\right]\Bigg\}.
\end{eqnarray}
Here, $t_{1}$ and $t_{2}$ denote the time of flight between source and
grating and between grating and detector, respectively, and
$\chi'_{t_{1}}(s,q)$ is the state immediately after the diffraction
grating. The density distribution on the screen is then given by
\begin{eqnarray}
w(x) & = &
\frac{1}{2\pi\hbar}\int_{-\infty}^\infty dq\:e^{-iqx/\hbar}\chi_{t_{1}+t_{
2}}(0,q)\nonumber
\\
 & = & \exp\left [\int_{0}^{t_{2}}d\tau\gamma(\tau+t_{1})\right ] \notag \\
& &\times \left[w_{0}(x)+\int_{-\infty}^\infty dx' \: w_{0}(x') h(x - x')\right],
\end{eqnarray}
where $w_{0}(x)$ is the interference pattern in absence of decoherence and
\begin{eqnarray}
&&h(x)=\frac{1}{2\pi\hbar}\int_{-\infty}^\infty dq\:e^{-iqx/\hbar} \hspace{3.5cm} \notag \\
&\times& \left[\exp\left\{\int_{0}^{t_{2}}d\tau\:\gamma(\tau+t_{1})\eta\left[\frac{q}{M}(t_{2}
-\tau)\right ] \right\}-1\right].
\end{eqnarray}
Assuming that the rate $\gamma(t)$ is constant in the chamber and zero outside yields
Eq.~\eqref{eq:signal}.

\end{appendix}

\bibliographystyle{klausbook}

\end{document}